\documentclass[aps,twocolumn,floatfix,amsmath,amssymb]{revtex4}
\usepackage{graphicx}
\usepackage{dcolumn}
\usepackage{bm}
\usepackage{epstopdf}

\begin{document}

\title{Highly resolved measurements of Stark-tuned F\"orster resonances between Rydberg atoms}
\author{J. Nipper}
%	\email{j.nipper@physik.uni-stuttgart.de}
\author{J.B. Balewski}
\author{A.T. Krupp}
\author{B. Butscher}
\author{R. L\"{o}w}
\author{T. Pfau}
\affiliation{
5. Physikalisches Institut, Universit\"{a}t Stuttgart, Pfaffenwaldring 57, 70569 Stuttgart, Germany.}

\date{\today}

\begin{abstract}
We report on experiments exploring Stark-tuned F\"orster resonances between Rydberg atoms with unprecedented resolution in the F\"orster defect. The individual resonances are expected to exhibit different angular dependencies, opening the possibility to tune not only the interaction strength but also the angular dependence of the pair state potentials by an external electric field.
We achieve a high resolution by optical Ramsey interferometry for Rydberg atoms combined with electric field pulses. The resonances are detected by a loss of visibility in the Ramsey fringes due to resonances in the interaction. We present measurements of the density dependence as well as of the coherence time at and close to F\"orster resonances.
\end{abstract}

\maketitle

Rydberg atoms in ultra cold atomic systems are particularly interesting for negligible motional dephasing ('frozen Rydberg gas'), strong interactions and various options to control them coherently. With this they are promising ingredients for quantum information processing \cite{Ja,Lu,SWM10} and quantum simulation \cite{WML10}. Also exotic phases for Rydberg dressed ensembles of atoms \cite{HNP10,PMB10,We} are proposed. These applications rely on coherent control of the strong interactions. Here we study the coherence in the presence of these interactions.\\
One possibility to control interactions between Rydberg atoms are so-called F\"orster resonances. Two dipole coupled pair states become degenerate and create a resonant dipole-dipole interaction between the atoms. As accidental degeneracy is unlikely, certain Rydberg states can be tuned into F\"orster resonance by microwave fields \cite{ABC06,BPM07} or a small electric field \cite{RCK07}. Different magnetic substates can be coupled by different polarizations of the coupling dipole. This generates diverse angular dependences for different F\"orster resonances. Thereby Stark tuned F\"orster resonances offer the possibility to control both, the interaction strength and the angular dependence by switching small electric fields. They have been studied in several seminal experiments in terms of dipole blockade \cite{Vo}, line shape analysis \cite{RT}, double-resonance spectroscopy \cite{RYL08} and excitation statistics \cite{RR}. Until now these experiments did not resolve the magnetic splitting of the F\"orster resonances.\\
In order to study coherent control of these interactions, interferometric methods offering phase sensitivity are well suited. As already pointed out by Ramsey in 1950 \cite{R50} interferometric schemes relying on separated oscillating fields are advantageous in many aspects compared to a single pulse of the coupling field. Besides an increased spectral resolution it allows to study coherent phenomena not being limited by spatial inhomogeneities of the coupling field. Ramsey interference methods were already used to investigate the coherence in resonant microwave coupling of single-atom Rydberg states \cite{RTB03} and in the coupling between pair states \cite{ARM02}. These experiments could not coherently control the excitation and could not study the decoherence directly at the F\"orster resonance. To our knowledge so far no experiment has been performed that coherently controls both the laser excitation and the interaction of Rydberg atoms.\\
Here, we apply optical Ramsey spectroscopy to coherently excite and de-excite $^{87}$Rb atoms to the 44d Rydberg state. 
\begin{figure}[t]
\begin{center}
\resizebox{0.93\columnwidth}{!}{\includegraphics{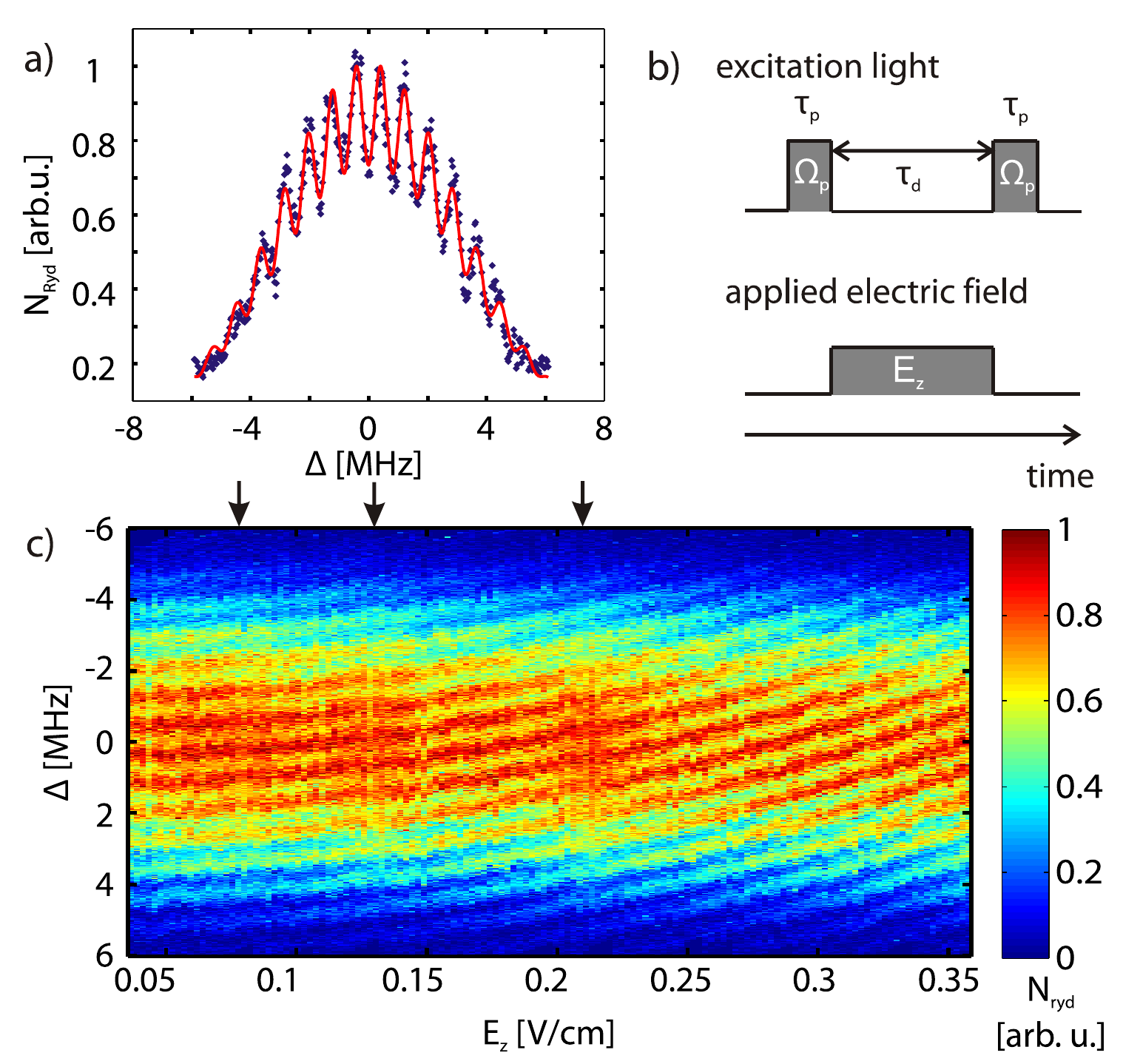}}
\caption{\label{fig1} (a) Single shot Ramsey spectrum for a pulsed electric field of $E_z$\,=\,0.3\,V/cm (blue dots) and least square fit to the data (red line). $\Delta$ is the detuning of the exciting laser to the atomic resonance. (b) Pulse sequence used throughout this paper. (c) Color coded Ramsey spectra for varying pulsed electric fields. With increasing electric field a phase shift of these fringes occurs that depends quadratically on the electric field. A loss in visibility at 0.08\,V/cm, 0.13\,V/cm and 0.21\,V/cm is visible, marked by the arrows in c). (a,c) Parameters for Ramsey measurements: $\tau_p$\,=\,0.15\,$\mu$s, $\tau_d$\,=1\,.0\,$\mu$s, $\rho$\,=\,$1.2\cdot10^{12}$\,cm$^{-3}$}
\end{center}
\end{figure}
These experiments can be viewed as an atom interferometer, similar to the atom-molecule interferometer in \cite{BNB10}. The phase of the two arms of the interferometer can be tuned independently by small electric fields. A full coherent control over the electronic state and the phase of the atoms is realized. Using this Ramsey spectroscopy we explore the dephasing at F\"orster resonances of the channel 
\begin{equation}
\label{foerster}
44d_{5/2}+44d_{5/2}\rightarrow 46p_{3/2}+42f.
\end{equation}
Several magnetic substates of the 42f-state, splitted by fine structure coupling, Stark and Zeeman effect, can contribute to this resonance. These substates can be tuned into F\"orster resonance at slightly different electric fields, reducing the fringe visibility due to interaction induced dephasing. This splitting of the F\"orster resonance is resolved in the measurements, as shown in Fig. \ref{fig1} and \ref{fig2}.\\
Experimentally $^{87}$Rb atoms are initially prepared in the $f$\,=\,2 $m_f$\,=\,2 ground state in a magnetic trap. After evaporative cooling the atom number is varied by a Landau-Zener sweep and the trap offset is adiabatically ramped to a magnetic field of $13.55$\,G. This is done to shift Stark-induced crossings of Zeeman split substates of single atom Rydberg states to electric fields out of the experimentally interesting range. Temperatures of about 1\,$\mu$K at densities between $\rho$\,=\,$1\cdot10^{10}$\,cm$^{-3}$ and $\rho$\,=$\,1.2\cdot10^{12}$\,cm$^{-3}$ are realized, above the critical temperature for Bose-Einstein condensation. Throughout this paper the atoms are coherently excited to the 44d$_{5/2}$, m$_j$\,=\,5/2 Rydberg state via a two-photon process, detuned by $2\pi\cdot400$\,MHz to the intermediate 5p$_{3/2}$-state. The total laser linewidth is below $2\pi\cdot100$\,kHz and the single atom two-photon Rabi frequency is $\Omega\approx2\pi\cdot100$\,kHz. Details about the experimental setup can be found in \cite{L07}. For Ramsey spectroscopy two short laser pulses of $\tau_p$\,=\,0.15\,$\mu$s duration, separated by a variable delay time $\tau_d$\,=\,$0\ldots2$\,$\mu$s, are applied to the atoms (Fig. \ref{fig1}b)). The Rydberg atom number $N_{Ryd}$ is measured after the second light pulse by field ionization and ion detection of all Rydberg states. The sequence of excitation and detection is repeated 401 times in one atomic sample so that one entire Ramsey spectrum is measured in one atomic cloud. No averaging over spectra from different atomic samples is necessary. Fig. \ref{fig1}a) shows such a single shot spectrum. The appearence of a Ramsey fringe pattern in frequency space proves the coherence of the excitation process.\\
Additionally a pulsed electric field $|\vec{E}|$ is switched on within 20\,ns during the entire delay time (Fig. \ref{fig1}b)). The electric field component $E_z$ along the long axis of the magnetic trap was calibrated by measuring the Stark effect of the 44d-state. Note that a small radial electric field, possibly on the order of $0.05$\,V/cm, can not be controlled in the experiment and contributes to $|\vec{E}|$.\\
Due to the high polarizability of Rydberg atoms the electric field detunes the Rydberg state relative to the exciting laser during the delay time when no excitation light is applied. This generates a phase shift 
\[\phi=-\frac{1}{\hbar}\int{\frac{\alpha}{2} |\vec{E(t)}|^2 dt}\]
between the Rydberg and the ground state atoms, where $\alpha$ is the polarizability of the 44d-state. A phase shift in the interference fringes appears and is experimentally apparent in the quadratic dependence of the fringe pattern on the electric field in Fig. \ref{fig1}c).
\begin{figure}[t]
\begin{center}
\resizebox{0.95\columnwidth}{!}{\includegraphics{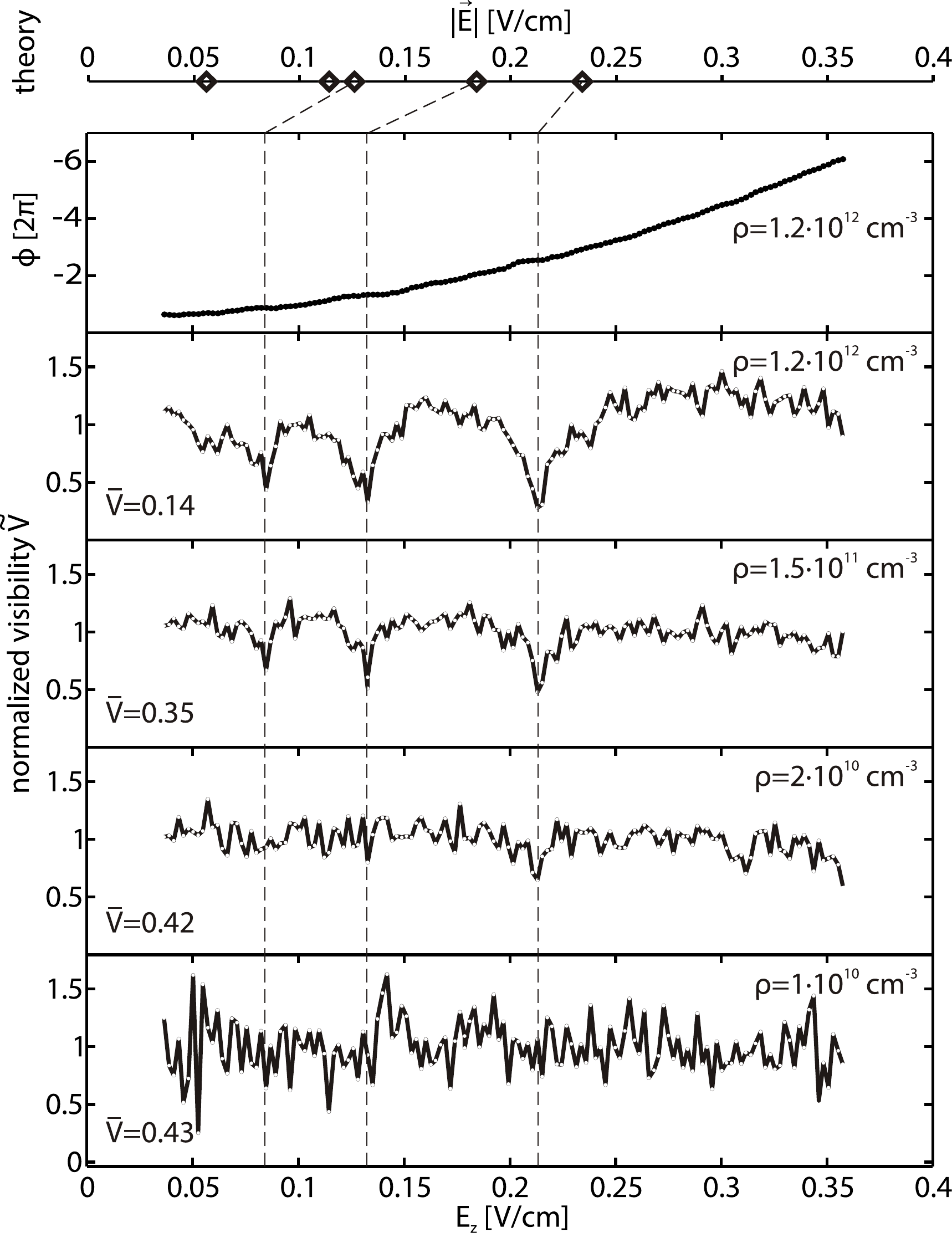}}
\caption{\label{fig2} The uppermost graph shows the phase $\phi$ of the ramsey fringes. The lower panels show the visibility normalized to the mean value $\bar{V}$ of each dataset against electric field for different densities $\rho$ of ground state atoms. The magnetic field is 13.55 G and the Rabi frequency is constant for all measurements. The lower electric field axis, valid for all graphs, denotes the calibrated component of the electric field, $E_z$. The dashed vertical lines indicate the resonant electric fields obtained from measurements. On the upper electric field axis (total electric field $|\vec{E}|$, including an experimentally not accessible radial component) the calculated resonant electric fields are indicated by diamonds.}
\end{center}
\end{figure} 
No loss in the visibility of the fringes is visible even for phase shifts of almost $12\pi$ at 0.35\,V/cm. This shows the remarkable stability of the coherence with respect to homogeneous fields and realizes complete coherent control over the state of the atoms.\\
This Ramsey interferometer is now used to study the F\"orster resonances. Here, the visibility $V$
\[V=\frac{\text{max}(N_{Ryd})-\text{min}(N_{Ryd})}{\text{max}(N_{Ryd})+\text{min}(N_{Ryd})}\]
provides an observable that is sensitive to decoherence and dephasing processes. It is obtained from a fit to each individual spectrum (see Fig. \ref{fig1}a)). Figure \ref{fig2} shows the normalized visibility $\tilde{V}$\,=\,$V/\bar{V}$, where $\bar{V}$ is the mean visibility of each dataset, for different densities of ground state atoms. For high densities distinct dips in the visibility can be seen at 0.08\,V/cm, 0.13\,V/cm and 0.21\,V/cm. For decreasing densities of ground state atoms these features diminish. For the lowest attainable density of $1\cdot10^{10}$\,cm$^{-3}$ the noise level is increased due to the weak signal. However, none of the dips are visible. The disappearence of the features for lower densities is a clear sign for an interaction process. We attribute these features to F\"orster resonance interaction.\\
The origin of the loss in visibility at the F\"orster resonances can already be qualitatively understood in a two-body picture. Dipole-dipole coupling of a pair of atoms in the $44d$-state to the $46p$ and $42f$-states during the electric field pulse will lead to a phase shift of the doubly excited state relative to a state where only one atom is excited. This leads to a dephasing within the system and is visible as a reduction of the visibility in the Ramsey fringes. Furthermore, an emerging population in the $p$ and $f$-states will not be coupled by the second light pulse of the Ramsey sequence, reducing the visibility in the Ramsey fringes even more.\\
However, since the dipole-dipole interactions depend on the interatomic distance as $1/r^3$, inhomogeneous Rydberg atom distributions in the experiment will lead to bands of interaction shifts for the collectively excited atoms \cite{CP10}. This results in an additional loss of visibility in the experiment. Furthermore, in our experimental conditions many-body effects are expected to contribute to the strength of the dephasing and the lineshape \cite{MCT98,YRP09} and have to be taken into account for a qualitative study. However, the resonance positions are not expected to be notably shifted by many-body phenomena.\\
Note that the loss of coherence due to the F\"orster interaction does not lead to a reduced resolution in the spectroscopy since in the experimental sequence the atoms are excited far off the F\"orster resonance.\\
To calculate the required fields for a F\"orster resonance we diagonalize the Hamiltonian
\[H=H_0+H_F+H_B,\]
where $H_0$ is the single atom field-free Hamiltonian, $H_F$\,=\,$\vec{d}\cdot\vec{E}$ is the electric field and $H_B$\,=\,$\vec{\mu}\cdot\vec{B}$ the magnetic field Hamiltonian. The electric and magnetic dipole matrix elements, $d_{ij}$ and $\mu_{ij}$ respectively, are calculated from numerical integrations of the Schr\"odinger equation, using the quantum defects from \cite{Li,Lo,HJN06}. These calculations are done following the approach of \cite{RCK07} but taking a magnetic field into account. From the eigenstates in the magnetic and electric field the crossings of the pair states of equation \ref{foerster} and the angular dependent interaction strength
\[U(\Theta)=\sqrt{2}\cdot\left<\left.42f\right|\right.\left<\left.46p\right|\right.V_{dd}(\Theta)\left.\left|44d\right>\right.\left.\left|44d\right>\right.\] 
of the resonances can be calculated. $V_{dd}(\Theta)$ is the dipole-dipole interaction operator and $\Theta$ the angle between the direction of the electric field and the interatomic axis. Compared to calculations without magnetic field we get additional splittings and more resonances. Figure \ref{figX} shows the dependence of the electric fields that tune the pair state potentials into resonance as a function of the magnetic field. Here, the magnetic and electric fields are parallel. Most of the resonant pair states are not coupled, as the involved single atom states, depending on their eigenstates in the magnetic and electric field, are not dipole coupled. Hence they do not induce interactions. The angular maximum $U_{max}$ of the strength of the interaction at resonance is indicated by the diameter and the color of the dots.
\begin{figure}[t]
\begin{center}
\resizebox{0.95\columnwidth}{!}{\includegraphics{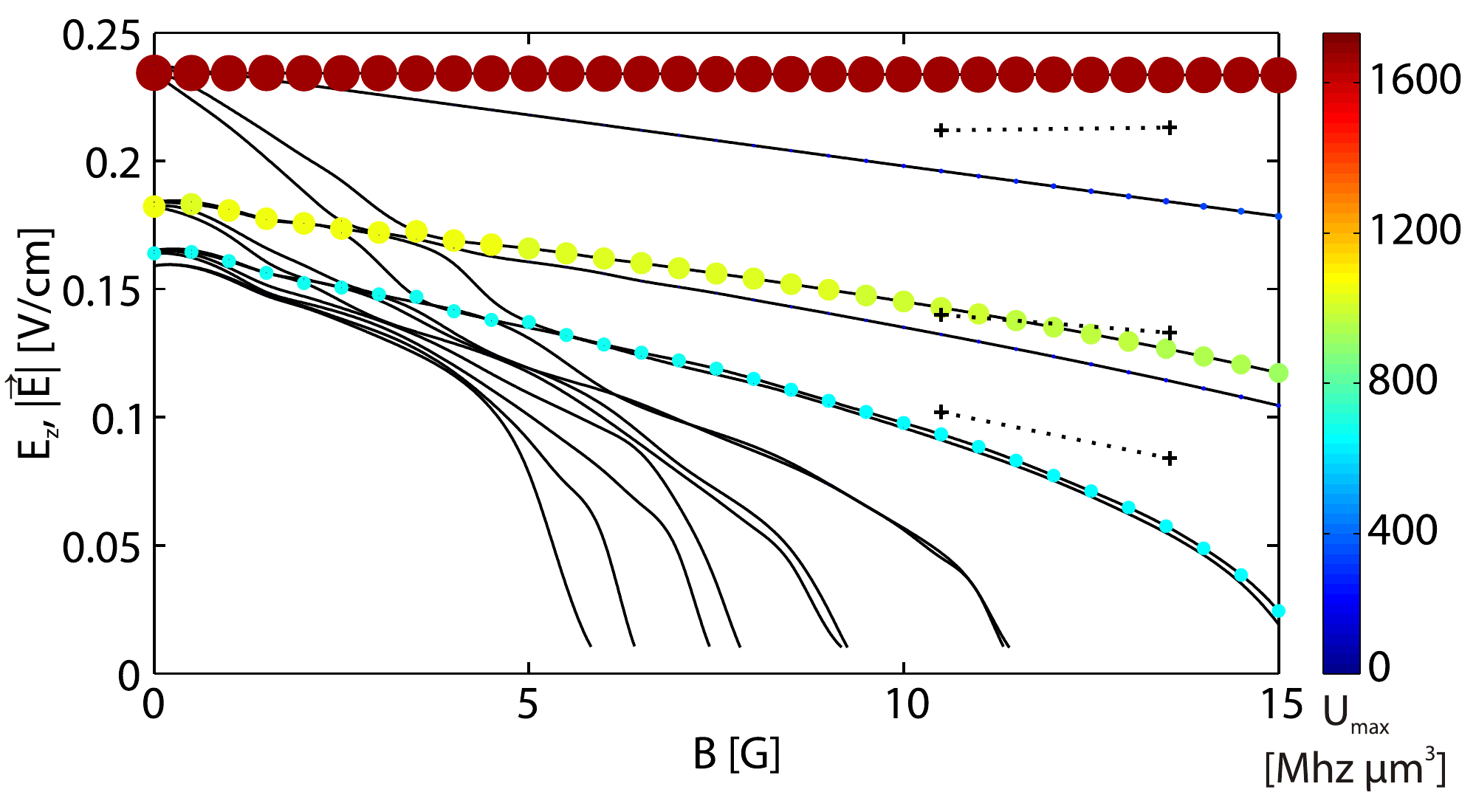}}
\caption{\label{figX} Magnetic field dependence of the electric fields $|\vec{E}|$ required to tune the pair states into F\"orster resonance for parallel electric and magnetic field (dots and lines). Additional splittings occur in finite magnetic field due to the splitting into magnetic sublevels. The angular maximum of the interaction strength $U_{max}$ of the resonances is indicated by the color and the diameter of the dots, that linearly increases with the strength of the resonance. The strongest resonance is at $|\vec{B}|$\,=\,0\,G and $|\vec{E}|$\,=\,0.234\,V/cm with a strength of 1556\,MHz$\cdot\mu$m$^3$. The crosses indicate the measured resonance positions in $E_z$ at 10.50\,G and 13.55\,G, i.e. uncorrected for the radial electric field.}
\end{center}
\end{figure}
In zero magnetic field three resonances exist that are dipole-dipole coupled. They differ in the involved magnetic substate of the 46$f$-state. In a basis formed of magnetic quantum numbers for electron spin and orbital angular momentum ($\left.\left|m_s,m_l\right>\right.$) the substates at $B$\,=\,0\,G can be identified as $\left.\left|\frac{1}{2},3\right>\right.$, $\left.\left|\frac{1}{2},2\right>\right.$ and $\left.\left|\frac{1}{2},1\right>\right.$ for the resonances at 0.234\,V/cm, 0.182\,V/cm and 0.164\,V/cm respectively. Small differences to \cite{RCK07} are due to the different quantum defects. In a finite magnetic and electric field the states mix and split in several substates. The resonance at $|\vec{E}|$\,=\,0.234\,V/cm is almost independent of the magnetic field.\\
In Fig. \ref{fig2} the calculated resonant electric fields at $13.55$\,G are indicated by diamonds. In the experiment only three resonances could be clearly identified. The resonance at $E_z$\,=\,0.21\,V/cm was observed to be not shifted by the magnetic field, as shown in Fig. \ref{figX}. Based on this magnetic field dependence the measured resonances were attributed to the calculations as indicated by the dashed lines in Fig. \ref{fig2}. Calculations show that the discrepancy can be explained by a reasonable, experimentally not controllable radial stray field on the order of 0.05\,V/cm. The radial field changes the relative angle between the magnetic and the electric field. That must be included in the calculations and leads to shifts of the resonances. Due to the stray field the calculated resonance at $|\vec{E}|$\,=\,0.06\,V/cm is not observable. The doublet at about $|\vec{E}|$\,=\,0.12\,V/cm can not clearly be separated in the measurement.\\
To study the emergence of decoherence in more detail the dephasing rates are obtained in the following: Ramsey spectra with varying delay times are taken for fixed electric fields $E_z$ and fixed atomic densities $\rho$. Figure \ref{fig34}a)
\begin{figure}
\begin{center}
\resizebox{0.9\columnwidth}{!}{\includegraphics{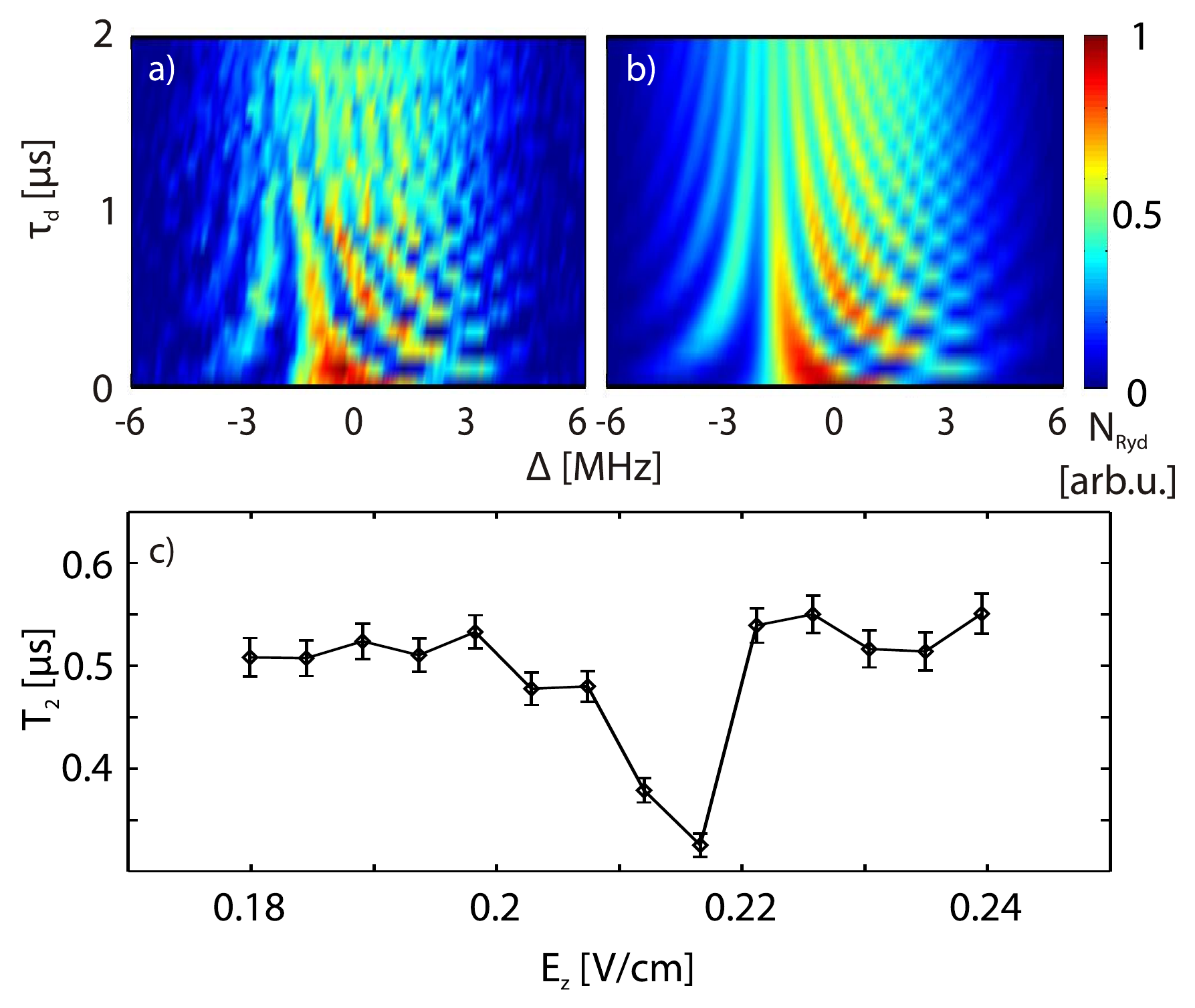}}
\caption{\label{fig34} Ramsey experiment with variable delay times between $\tau_d$\,=\,0\,$\mu$s and $\tau_d$\,=\,2\,$\mu$s at an atom density of $1.2\cdot10^{12}$\,cm$^{-3}$. a) Color coded experimental data and b) data obtained by a numerical fit, both at $E_z$\,=\,0.21\,V/cm. c) Dephasing time $T_2$ obtained from the numerical fits for different electric fields $E_z$. The errorbars denote the standard deviation of the fit parameter.}
\end{center}
\end{figure}
 shows such a set of data where the delay time and with it the length of the electric field pulse is varied between $0\,\mu$s and $2\,\mu$s. For longer delay times the fringe frequency is higher, as expected for a Ramsey experiment.\\
A numerical solution of the optical Bloch equations \cite{AE}
\begin{align}
	\dot{u}=&-\Delta v-\frac{u}{T_2}\nonumber\\
	\dot{v}=&\Delta u+\Omega w-\frac{v}{T_2}\nonumber\\
	\dot{w}=&-\Omega v-\frac{w+1}{T_1}\nonumber
\end{align}is fitted to the experimental data, where $T_1$ is the excited state lifetime, $T_2$ the dephasing time accounting for all energy conserving dephasing processes, $\Delta$ is the detuning and $\Omega$ the two-photon Rabi frequency. As $T_1$ is much longer than the duration of each sequence it is fixed to $T_1$\,=\,47\,$\mu$s, the calculated lifetime for the 44d$_{5/2}$-state \cite{BRT09}. Fits with $T_1$\,=\,100\,$\mu$s and $T_1$\,=\,25\,$\mu$s show that the results for $T_2$ vary only within the standard deviation of the fit. The remaining two fit parameters are the dephasing time $T_2$ and a numerical factor $N$, proportional to the Rydberg atom number. These measurements are repeated for different electric fields and the obtained electric field dependence of the dephasing time is plotted in Fig. \ref{fig34}c).
A reduced dephasing time is visible for an electric field of $0.214$\,V/cm, identical to the observation from Fig. \ref{fig1}c). Note that this dephasing time does not only result from binary interactions, but also includes effects like many-body interactions and a finite magnetic field broadened linewidth.\\
%Offresonant to this F\"orster resonance the 44d-state also exhibits severe interactions that originate in the small F\"orster defect \cite{RCK07}. These interactions possibly reduce the dephasing time compared to a non-interacting state. On top of this resonant dipole-dipole interaction occurs at the resonant electric field $E_z$\,=\,0.214\,V/cm leading to the even shorter dephasing time.\\
The measurement of T$_2$ is not limited by technical constraints as the intrinsic dephasing in the experiment, dominated by the laser linewidth, is substantially lower than the measured dephasing. Measurements on rather weakly interacting Rydberg molecules \cite{BNB10} show that considerably longer dephasing times can be measured.\\
In conclusion we performed coherent Ramsey spectroscopy with Rydberg atoms. We realized a Stark-tuned phase shifter to measure single atom and ensemble properties. Using this Ramsey spectroscopy we resolve several F\"orster resonances. From the width of the signal at $0.21$\,V/cm of $\sim0.01$\,V/cm a resolution of $\sim5$\,MHz for the F\"orster defect can be calculated from the differential Stark shifts of the pair states. This is clearly below the splitting of the resonances and opens the possibility to tune the angular dependence of the interaction. So far the resolution was on the same order as the splitting \cite{BPM07}. Furthermore coherence times for atoms in the 44d-state at and near the Stark-tuned F\"orster resonance were obtained. The next steps will involve a detailed study of the origin of the dephasing process. Dephasing measurements on the resonant energy transfer process \cite{ARM02} will be performed in future measurements. In order to reduce the dephasing due to an inhomogeneous arrangement of atoms a more ordered system like an optical lattice or several small dipole traps could be used, where possibly interactions induce a collective phase shift of the atoms. Control over the angular dependence of the interaction will create an anisotropy in the Rydberg blockade and accordingly an anisotropic Rydberg density distribution. The Ramsey interferometer in combination with tuneable strong two-body interactions is a well suited tool to study and control this angular dependence of the interactions.

\begin{acknowledgments}
This work is funded by the Deutsche Forschungsgemeinschaft (DFG) within the SFB/TRR21 and the project PF~381/4-2. We also acknowledge support by the ERC under contract number 267100.
\end{acknowledgments}

\end{document}